\documentclass[twocolumn,english,twocolumn,english,aps,pra,nofootinbib]{revtex4-1}
\usepackage{lmodern}
\usepackage[T1]{fontenc}
\usepackage[utf8]{inputenc}
\usepackage{textcomp}
\usepackage{mathrsfs}
\usepackage{amsmath}
\usepackage{amssymb}
\usepackage{stackrel}
\usepackage{graphicx}
\usepackage{babel}
\begin{document}
\title{Zero-field magnetometry based on the combination of atomic orientation
and alignment}
\author{Gwenael Le Gal$^{1,2}$}
\email{gwenael.legal@yahoo.fr}

\author{Agustin Palacios-Laloy$^{1}$}
\address{1. Univ. Grenoble Alpes, CEA, Leti, F-38000 Grenoble, France}
\address{2. Univ. Grenoble Alpes, CNRS, Grenoble INP, G2Elab, F-38000 Grenoble,
France}
\begin{abstract}
Optically pumped magnetometers usually rely on optical pumping using
circularly- or linearly-polarized light. We study here zero-field
magnetometers pumped with elliptically-polarized light, preparing
both atomic orientation and alignment with complementary geometries.
We start by extending the ``three-step approach'' for elliptically-polarized
pumping. This allows us studying the Hanle effect in elliptical polarization
by comparing the analytical absorption signals with experiments made
on helium-4 metastable state. We then study parametric resonance magnetometers
based on elliptical polarization by using the dressed-atom formalism
with one and two radio-frequency fields. The results show a good agreement
with the experimental measurements and open interesting perspectives
for magnetometry where symmetry breaking by pumping light is mitigated.
\end{abstract}
\maketitle

\section{Introduction}

During the last years, optically pumped magnetometers (OPMs) operating
in very low magnetic fields have reached excellent levels of sensitivity
surpassing those of SQUIDs \citep{kominis_subfemtotesla_2003,vasilakis_generation_2015,shah_high_2010}
without requiring cryogeny. Such sensors already proved their ability
to measure ultra-low magnetic fields in several domains such as space
exploration \citep{slocum_advances_1970} or magnetic imaging of biological
currents in cardiography \citep{weis_mapping_2010,morales_magnetocardiography_2017},
fetal cardiography \citep{wyllie_optical_2012} and encephalography
\citep{xia_magnetoencephalography_2006,labyt_magnetoencephalography_2019}.

Most of the vector zero-field OPMs configurations use circularly-polarized
light \citep{cohen-tannoudji_diverses_1970-1,jimenez-martinez_optically_2014,shah_fully_2018}
for pumping the atomic ensemble towards an oriented state, i.e. with
average angular momentum $\left\langle J_{k}\right\rangle \neq0$
where $\overrightarrow{k}$ is the propagation direction of the light.
For atomic states with total angular momentum $J\geq1$, such as the
$2^{3}\mathrm{S}_{1}$ helium-4 ($^{4}$He) metastable state, one
can use linearly-polarized light to prepare atomic alignment (i.e.
states with $\left\langle 3J_{e}^{2}-\overrightarrow{J}^{2}\right\rangle \neq0$
where $\overrightarrow{e}$ is the direction of the pump-light electric
field $\overrightarrow{E_{0}}$ \citep{shi_cesium_2018}). In both
cases, the symmetry breaking by the optical pumping prevents from
measuring the magnetic field component longitudinal to the pumping
direction ($\overrightarrow{k}$ for orientation, $\overrightarrow{e}$
for alignment) with a good sensitivity.

Since elliptically-polarized light has both circular and linear polarization
components, it can be used for pumping the atomic ensemble towards
states that are both oriented and aligned. In a recent communication
\citep{le_gal_parametric_2021}, this kind of polarization combined
to parametric resonances \citep{dupont-roc_etude_1971,beato_theory_2018}
resulting from two radio-frequency (RF) magnetic fields allowed measuring
the three components of the magnetic field with isotropic sensitivity.

Our purpose here is to provide an in depth study of the physics of
zero-field magnetometers based on elliptically-polarized light.

To do so, we focus first (in Sec. \ref{sec:Hanle}) on the Hanle effect
of an ensemble pumped using elliptically-polarized light, in analogy
to the studies of Castagna and Weis \citep{castagna_measurement_2011}
for orientation and Breschi and Weis \citep{breschi_ground-state_2012}
for alignment. For elliptical pumping light Hanle resonances exist
for all components of the magnetic field. The amplitudes and widths
of these resonances depend on the pumping light ellipticity, which
allows an interesting analysis of the relaxation rates of the different
atomic multipoles.

From this study we deduce (in Sec. \ref{sec:Single-RF}) a theoretical
description of a single-RF field parametric resonance magnetometer
(PRM) based on elliptically-polarized light by using the dressed atom
formalism \citep{dupont-roc_etude_1971,beato_theory_2018,bevilacqua_harmonic_2020}.

The dressed atom formalism also allows studying an atomic ensemble
subject to several RF fields \citep{dupont-roc_etude_1971,beato_theory_2018},
a configuration which allows the measurement of several components
of the magnetic field. We present these calculations in Sec. \ref{sec:Double-RF}.
These predictions are in good agreement with the experimental measurements.
We will also discuss the choices of parameters---ellipticity, RF
fields directions, frequencies and amplitudes---which are optimal
for obtaining a three-axis vector magnetometer with isotropic sensitivity.

\section{Hanle effect of a spin-1 atomic state pumped with elliptically-polarized
light}

\label{sec:Hanle}

\subsection{Theory}

\label{subsec:Theory} Hanle effect is a well-known phenomenon, which
consists in resonant variations of the optical properties of a polarized
atomic ensemble as a function of the magnetic field \citep{hanle_uber_1924}.
Such resonances only appear when the magnetic field $B_{0}$ is very
small ($\gamma B_{0}\ll\Gamma$, where $\gamma$ is the gyromagnetic
ratio and $\Gamma$ is the relaxation rate of the Zeeman coherences,
$-2\pi\times28\:\mathrm{s^{-1}.nT^{-1}}$ and $\sim5\:\mathrm{ms^{-1}}$
respectively for the $2^{3}\mathrm{S}_{1}$ state of helium-4), and
when the atomic polarization is transverse to it \citep{cohen-tannoudji_detection_1969,brazhnikov_shift_2020}.
Our goal here is to calculate the absorption signals resulting from
the Hanle effect for any polarization of the pumping light, and with
respect to all the components of the magnetic field.

\begin{figure*}[t]
\centering

\includegraphics[scale=0.92]{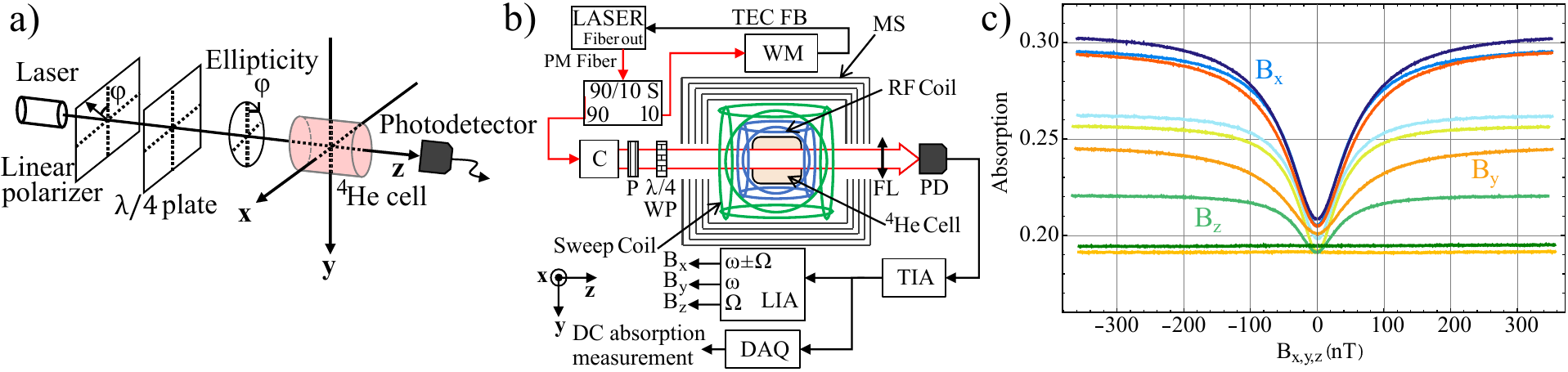}

\caption{\label{fig:1}Theoretical and experimental study of Hanle effect resonances
in $^{4}$He atoms pumped with elliptically-polarized light. (a) Geometrical
configuration considered for the optical pumping of the $^{4}$He
ensemble with elliptically-polarized light. The light goes through
a linear polarizer, forming an angle $\varphi$ with the $\protect\overrightarrow{y}$
axis, and a quarter waveplate with fast axis parallel to $\protect\overrightarrow{y}$
before entering the $^{4}$He cell. (b) Experimental setup. PM: Polarization
maintaining; TEC FB: TEC Feedback; WM: Wavelength Meter; MS: Magnetic
Shield; 90/10 S: 90/10 Splitter; C: Collimator; P: Linear Polarizer;
WP: Waveplate; FL: Focusing Lens; PD: InGaAs Photodiode; TIA: Transimpedance
Amplifier; LIA: Lock-In Amplifier; DAQ: DAQmx Board. The redlined
paths show the optical paths and the black ones the electrical signal
paths. (c) Hanle effect resonances observed experimentally when sweeping
$B_{x}$ (blue), $B_{y}$ (orange) and $B_{z}$ (green) at different
ellipticities: $\varphi=0{^\circ}$ (lighter colors), $\varphi=25{^\circ}$
(middle shade colors) and $\varphi=45{^\circ}$ (darker colors). See
the text for the description of the experiment.}
\end{figure*}

We consider the setup of Fig. \ref{fig:1}.a, in which an ensemble
of $^{4}$He atoms in the metastable state are subject to optical
pumping using elliptically-polarized light tuned on the $\mathrm{D}_{0}$
transition ($2^{3}\mathrm{S}_{1}\rightarrow2^{3}\mathrm{P}_{0}$).
In order to describe the atomic polarization, we decompose the metastable
state density matrix $\rho$ of the ensemble on the irreducible tensor
operators (ITO) basis:

\begin{equation}
\rho=\stackrel[k=0]{2J}{\sum}\:\stackrel[q=-k]{k}{\sum}m_{q}^{(k)}\hat{T}_{q}^{(k)\dagger}\label{eq:rhoITO}
\end{equation}

\noindent with $m_{q}^{(k)}=\left\langle \hat{T}_{q}^{(k)}\right\rangle $
the atomic multipole moments, $\hat{T}_{q}^{(k)}$ the irreducible
tensors operators and $J=1$ for the $2^{3}\mathrm{S}_{1}$ state.
The rank $k=0$ describes the total state population, rank $k=1$
the atomic orientation, and rank $k=2$ the atomic alignment. In the
following, we set the quantization axis along the light propagation
direction $\overrightarrow{z}$.

First, we want to derive a simple expression of the absorption signals.
For this purpose, we use the so-called three step approach, broadly
used in atomic magnetometry, which consists in modeling the dynamics
of the atomic polarization as it was happening in three steps: (i)
atomic state preparation by optical pumping, (ii) state evolution
under magnetic field and relaxation, and (iii) measurement of the
system state. This simplification allows obtaining a good picture
of the ensemble evolution as far as the pumping-light intensity is
low enough \citep{weis_weis_1993,kanorsky_quantitative_1993,budker_resonant_2002},
i.e. $\Gamma_{p}\ll\Gamma_{e}$, where $\Gamma_{e}$ is the relaxation
rate of the metastable state due to collisions with the cell walls
and other species in the plasma, and $\Gamma_{p}$ is the optical
pumping rate as defined in the references \citep{barrat_etude_1961,nacher_optical_1985,beato_second-order_2020}.

We will also make a few more reasonable approximations. The $^{4}$He
metastable state is populated by a high-frequency electrical discharge
(Sec. \ref{subsec:ExperimentalHanle}). We assume that the discharge
has reached a steady-state so that the metastable population is constant.
We also assume that the population of the $2^{3}\mathrm{P}_{0}$ is
negligible as compared to the one of $2^{3}\mathrm{S}_{1}$, since
the relaxation rate of the former is much larger than the one of the
latter.

Within those approximations, the dynamics of the metastable state
can be described by the following equation:

\begin{equation}
\left[\frac{d}{dt}-\mathbb{H}(\overrightarrow{B})+\Gamma\right]M=\Gamma_{p}M_{p}\label{eq:3step}
\end{equation}

\noindent for ranks $k=1$ and $k=2$ \citep{weis_theory_2006,breschi_ground-state_2012,beato_theory_2018,le_gal_dual-axis_2019}.
In this equation, $\Gamma=\Gamma_{e}+\Gamma_{p}$, $M$ is the multipole
moments tensor, i.e. the orientation vector $(m_{-1}^{(1)},m_{0}^{(1)},m_{1}^{(1)})^{t}$
for $k=1$, and the five-components alignment column matrix $(m_{-2}^{(2)},m_{-1}^{(2)},m_{0}^{(2)},m_{1}^{(2)},m_{2}^{(2)})^{t}$
for $k=2$. $\mathbb{H}(\overrightarrow{B})$ is the magnetic evolution
matrix, which for $k=1$ is given in Appendix \ref{sec:AppA} and
for $k=2$ is given in the references \citep{beato_theory_2018,le_gal_dual-axis_2019}.
$M_{p}$---with components $m_{q,p}^{(k)}$---is the steady-state
multipole moments tensor resulting from optical pumping alone, in
the absence of magnetic field and relaxation. For rank $k=0$ we have
constant population $m_{0}^{(0)}=1/\sqrt{2J+1}=1/\sqrt{3}$ \citep{blum_density_2012}.

In the usual cases when pumping with light is purely circularly- or
linearly-polarized the expressions of $M_{p}$ are well known \citep{seltzer_developments_2008,beato_theory_2018,le_gal_dual-axis_2019}\citep[Eq. 4.61 and 4.62]{blum_density_2012}.
For elliptical polarization, the expression of $M_{p}$ needs to be
carefully derived as a function of the light ellipticity.

With the setup shown in Fig. \ref{fig:1}.a: the resulting elliptically-polarized
light has its major axis along $\overrightarrow{y}$ and ellipticity
$\varphi$ ($\alpha=0$ and $\varepsilon=\varphi$ in the so-called
$\alpha-\varepsilon$ parametrization \citep{auzinsh_optically_2010}).
Following Omont \citep[Eq. 3.1]{omont_irreducible_1977} this leads
to the following non-zero components of $M_{p}$:\footnote{Note that our definition of the steady-state multipole moments is
not the same as in Eq. 4 of Beato \citep{beato_second-order_2020},
both being related by $M_{ss}=2M_{p}$. We made this choice in order
to keep the usual physical meaning of the pumping steady-state (otherwise
its corresponding density matrix has negative components). This choice
requires rewriting the Eq. 4 of Beato as $dM/dt=\mathbb{H}(\overrightarrow{B})\cdot M-\mathbb{R}\cdot M+2\Gamma_{p}M_{p}$.}

\begin{equation}
\begin{array}{c}
m_{0,p}^{(1)}=\dfrac{1}{2\sqrt{2}}\sin(2\varphi)\\
\\
m_{0,p}^{(2)}=-\dfrac{1}{2\sqrt{6}}\\
\\
m_{\pm2,p}^{(2)}=\dfrac{1}{4}\cos(2\varphi).
\end{array}\label{eq:Mpphi}
\end{equation}

\noindent When the light is strictly speaking elliptically-polarized
($\varphi\neq0{^\circ},45{^\circ}$) both a longitudinal orientation
and a transverse alignment are created in the atomic gas, the latter
along the ellipse major axis. The complete steady-state solutions
of Eq. \ref{eq:3step} using Eq. \ref{eq:Mpphi} are given in Appendix
\ref{sec:AppB}.

We can now calculate the photodetection signals. For the setup of
Fig. \ref{fig:1}.a and an optically thin ensemble, the absorption
coefficient $\kappa$ is \citep{laloe_relations_1969-1}:

\begin{multline}
\kappa\propto2\left(\frac{m_{0}^{(0)}}{\sqrt{3}}+\frac{m_{0}^{(2)}}{\sqrt{6}}\right)-2\sin(2\varphi)\frac{m_{0}^{(1)}}{\sqrt{2}}\\
-2\cos(2\varphi)\left(\frac{m_{-2}^{(2)}+m_{2}^{(2)}}{2}\right).\label{eq:kappaphi}
\end{multline}

The state population $m_{0}^{(0)}$ and the longitudinal alignment
$m_{0}^{(2)}$ always contribute to $\kappa$. Otherwise depending
on the relative strength between circular and linear polarization
the signal may be dominated by the longitudinal orientation $m_{0}^{(1)}$
or the transverse alignment $m_{\pm2}^{(2)}$.

For the Hanle effect (Eqs. \ref{eq:mkqHanleBz}, \ref{eq:mkqHanleBy}
and \ref{eq:mkqHanleBx}) the absorption coefficients are:

\begin{widetext}

\begin{equation}
\kappa_{Hanle}(\omega_{z},\omega_{x,y}=0)\propto\frac{2(\Gamma-\Gamma_{p})(\Gamma^{2}+4\omega_{z}^{2})+6\Gamma_{p}\omega_{z}^{2}\cos^{2}(2\varphi)}{3\Gamma(\Gamma^{2}+4\omega_{z}^{2})}\label{eq:kappaHanleBz}
\end{equation}

\begin{equation}
\kappa_{Hanle}(\omega_{x},\omega_{z,y}=0)\propto\dfrac{2}{\Delta_{x}}\left\{ (\Gamma-\Gamma_{p})(\Gamma^{4}+5\Gamma^{2}\omega_{x}^{2}+4\omega_{x}^{4})+\Gamma_{p}\cos^{2}(\varphi)\left[3\Gamma^{2}\omega_{x}^{2}+3\omega_{x}^{4}(1+3\sin^{2}(\varphi))\right]\right\} \label{eq:kappaHanleBx}
\end{equation}

\begin{equation}
\kappa_{Hanle}(\omega_{y},\omega_{z,x}=0)\propto\dfrac{2}{\Delta_{y}}\left\{ (\Gamma-\Gamma_{p})(\Gamma^{4}+5\Gamma^{2}\omega_{y}^{2}+4\omega_{y}^{4})+\Gamma_{p}\sin^{2}(\varphi)\left[3\Gamma^{2}\omega_{y}^{2}+3\omega_{y}^{4}(1+3\cos^{2}(\varphi))\right]\right\} \label{eq:kappaHanleBy}
\end{equation}

\end{widetext}

\noindent where $\Delta_{i}=3\Gamma(\Gamma^{4}+5\Gamma^{2}\omega_{i}^{2}+4\omega_{i}^{4})$,
with $i\in\{x,y\}$, $\omega_{x,y,z}=-\gamma B_{x,y,z}$ is the Larmor
angular frequency associated with the $x,y$ or $z$ component of
the magnetic field.

For every components of the magnetic field, there is an even-symmetric
absorption signals. The only exceptions are $\kappa_{Hanle}(\omega_{z},\omega_{x,y}=0)$
that is constant with $\varphi=45{^\circ}$, and $\kappa_{Hanle}(\omega_{y},\omega_{z,x}=0)$
which is constant when $\varphi=0{^\circ}$.

First, we have compared these expressions to the experimental measurements
of Hanle effect resonances (Sec. \ref{subsec:ExperimentalHanle}).
This allowed some interesting observations on the relaxation rates.
Secondly, we can use these Hanle effect dynamics as a basis to study
the parametric resonance signals, since parametric resonance can be
understood as the Hanle effect of the atom dressed by the RF fields
\citep{dupont-roc_etude_1971}.

\subsection{Experimental study of the Hanle effect}

\label{subsec:ExperimentalHanle}The experimental setup is shown on
Fig. \ref{fig:1}.b. It consists of a 1-cm diameter and 1-cm length
cylindrical cell filled with 9-Torr high purity helium-4. The $2^{3}\mathrm{S}_{1}$
metastable level is populated using a high-frequency (HF) capacitively-coupled
electric discharge at 13.23 MHz, absorbing 27 mW of electrical power.
We use an external cavity diode laser (Sacher Cheetah TEC 50), constantly
tuned to the $\mathrm{D_{0}}$ line of $^{4}$He, at $\lambda=1083.206$
nm, by locking its temperature with a wavelength-meter (HighFinesse
WS-7). The laser light is coupled in a polarization maintaining optical
fiber and passes through a variable optical attenuator before being
collimated using a converging lens to obtain a 7-mm diameter beam.
A linear polarizer and a zero-order quarter wave plate (Thorlabs reference
WPQ10M-1064), both mounted in independent rotation mounts, are placed
before the helium cell to control the pumping light ellipticity. In
the experiments reported here, the quarter wave plate fast-axis is
set along $\overrightarrow{y}$ and only the polarizer is rotated.

\begin{figure*}
\centering

\includegraphics[scale=0.86]{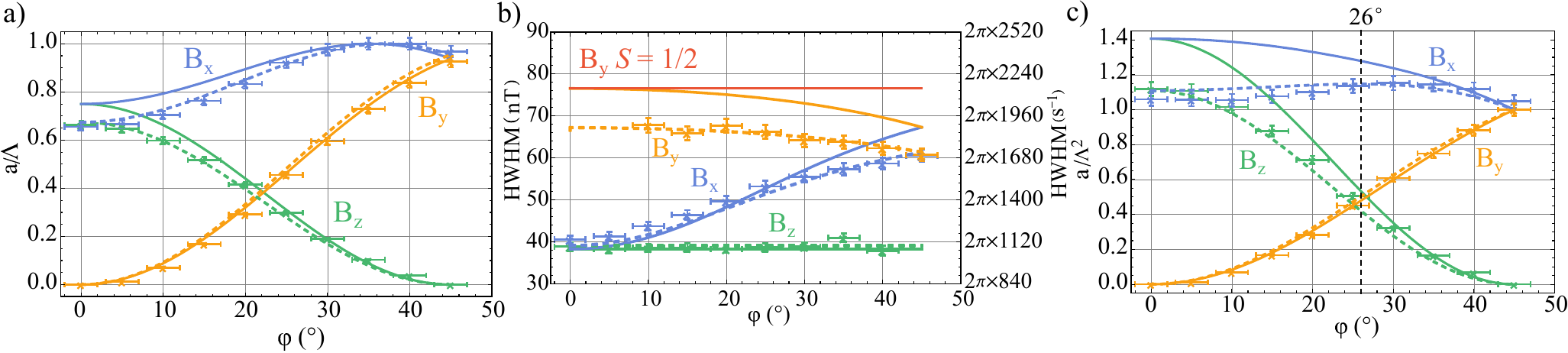}

\caption{\label{fig:4}Study of the Hanle effect with elliptically-polarized
pumping light. (a) Experimental (dots with error bars) and theoretical
(solid and dashed lines) dependences of the amplitude of the Hanle
resonances as a function of the pumping light ellipticity for $B_{x}$
(blue), $B_{y}$ (orange) and\textbf{ $B_{z}$} (green). The experimental
and theoretical data are normalized to the maximum amplitude. (b)
Experimental and theoretical HWHMs of the Hanle resonances as a function
of the light ellipticity. Red solid line: HWHM value of a spin-1/2
Hanle effect resonance at $\varphi=45{^\circ}$. (c) Experimental
and theoretical dependences of the PRM slope estimation $a/\Lambda^{2}$
as a function of the light ellipticity. The experimental data and
theoretical estimations are normalized to the value at $\varphi=45{^\circ}$
for $B_{y}$. The vertical black dashed line is set at the optimal
ellipticity to measure the three component of the magnetic field simultaneously.
For the three figures, the solid lines show the theoretical values
of the model with isotropic $\Gamma$, with $\Gamma_{e}=53.3$ nT
and $\Gamma_{p}=23.3$ nT. The dashed lines show the theoretical estimations
with anisotropic $\Gamma^{(k)}$, calculated with $\Gamma_{p}=23.3$
nT, $\Gamma_{e}^{(1)}=43.9$ nT and $\Gamma_{e}^{(2)}=54.9$ nT.}
\end{figure*}

The helium cell is placed inside two sets of triaxial Helmholtz coils:
the inner one is used to generate the RF fields when exciting parametric
resonances and the outer one is used to generate static magnetic field
sweeps. The cell and coils are placed inside a five-layer \textmu -metal
cylindrical magnetic shield, whose longitudinal axis is along $\overrightarrow{z}$.
A converging lens focuses the transmitted laser beam on an In-Ga-As
photodiode connected to a homemade transimpedance amplifier (TIA)
with gain $23.8\:\mathrm{k\Omega}$. Its output signal is acquired
by a NI-DAQmx board for Hanle effect measurements, or demodulated
with a Zürich MFLI lock-in amplifier for parametric resonances (see
Sec. \ref{sec:Single-RF} and \ref{sec:Double-RF}). For Hanle measurements,
a first-order low-pass filter with $40$ Hz cut-off frequency is inserted
before the DAQmx board to attenuate the noise brought by the plasma
discharge.

The three components of the magnetic field are sequentially swept
with ramps of $\pm300$ nT at 1 Hz frequency. The optical power is
set to $\sim300$ \textmu W at the cell input. The absorption is obtained
as $1-V_{\mathrm{PD}}/V_{\mathrm{PD,OFF}}$, where $V_{\mathrm{PD}}$
is the voltage at the TIA output during the magnetic field sweep,
and $V_{\mathrm{PD,OFF}}$ is the voltage when the helium-4 discharge
is off, which is measured before each acquisition.

Hanle effect signals for $\varphi$ ranging from $0{^\circ}$ (linear
polarization along $\overrightarrow{y}$) to $45{^\circ}$ (circular
polarization), and with respect to the three components of the magnetic
field are shown in Fig. \ref{fig:1}.c.

For every strictly speaking elliptical polarizations ($\varphi\neq0{^\circ},45{^\circ}$),
Hanle resonances can be observed for all the three component of the
magnetic field (e.g. middle shade colored lines in Fig. \ref{fig:1}.c
for $\varphi=25{^\circ}$).

We fit these curves with a Lorentzian function to obtain the amplitudes
and Half-Width-Half-Maximum (HWHM), noted $a/\Lambda$ and $\Lambda$
respectively. The results are shown in Fig. \ref{fig:4}.a and b,
along with the theoretical predictions computed from Eqs. \ref{eq:kappaHanleBz},
\ref{eq:kappaHanleBx} and \ref{eq:kappaHanleBy} (solid lines). The
theoretical HWHMs result from $\Gamma=\Gamma_{e}+\Gamma_{p}$ and
Eqs. \ref{eq:WidthHanle}. $\Gamma_{e}$ is estimated from the zero-field
parametric resonance versus $B_{z}$ with $\varphi=0{^\circ}$ at
low optical power ($P<13$ \textmu W), $\mathrm{HWHM}=\Gamma/2$.
If $\Gamma_{p}\ll\Gamma_{e}$, the HWHM is close to $\Gamma_{e}/2$.
$\Gamma_{p}$ is estimated as ($\mathrm{HWHM}-\Gamma_{e}$) at the
optical power $P\approx300$ \textmu W used in the measurements.

\begin{figure}
\includegraphics[scale=0.7]{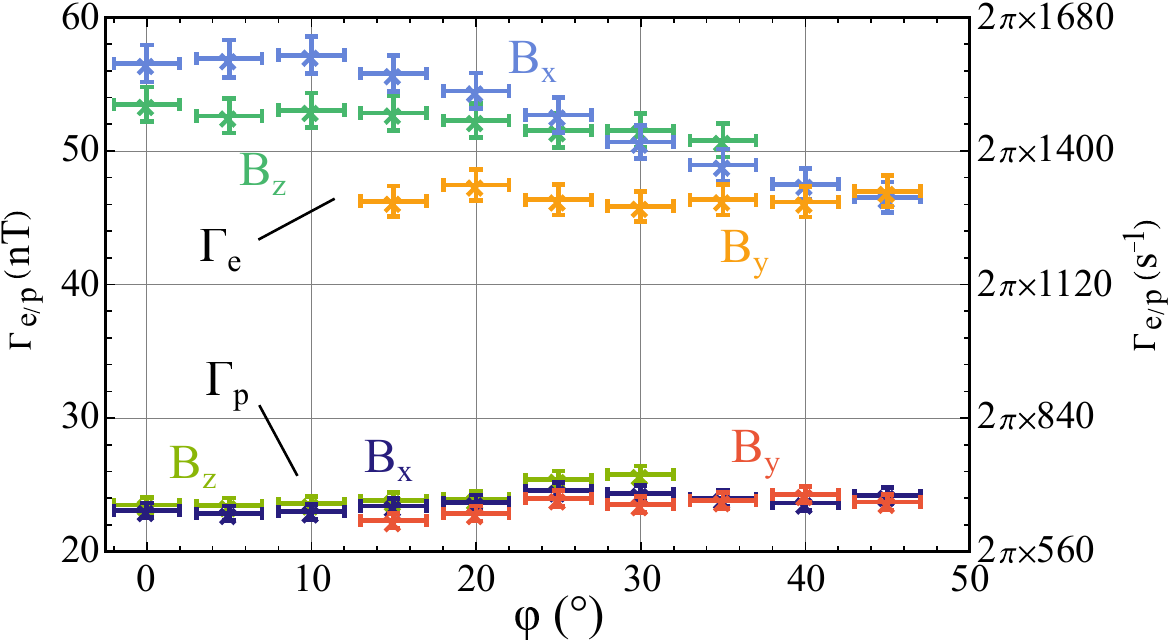}

\caption{\label{fig:5}Dependence of the fitted natural relaxation rate $\Gamma_{e}$
and the optical pumping rate $\Gamma_{p}$ with the pumping light
ellipticity $\varphi$. The values are obtained from the HWHM and
amplitude measurements of Figs. \ref{fig:4}.a and \ref{fig:4}.b,
as explained in Appendix \ref{sec:AppD}.}

\end{figure}

There is a qualitative agreement between the theoretical expectations
and the experiments. In Fig. \ref{fig:4}.a, we see that the $B_{x}$
resonance amplitude (orthogonal to the laser propagation direction
$\overrightarrow{z}$ and the quarter-wave plate fast axis $\overrightarrow{y}$)
increases with $\varphi$, and slightly decreases for $\varphi>40{^\circ}$.
The $B_{z}$ and $B_{y}$ resonances amplitudes show opposed behaviors
as a function of $\varphi$:
\begin{itemize}
\item The $B_{z}$ resonance amplitude decreases as $\varphi$ increases.
Hanle effect resonances being only present when the applied magnetic
field is transverse to the atomic polarization direction \citep{cohen-tannoudji_diverses_1970,dupont-roc_etude_1971,le_gal_dual-axis_2019},
this behaviour seems natural for this resonance, linked to the alignment
longitudinal to $\overrightarrow{y}$.
\item The $B_{y}$ resonance amplitude evolves in an opposite way, reaching
a higher relative amplitude than the alignment Hanle resonance. We
see in Fig. \ref{fig:4}.b that the HWHM also varies with $\varphi$,
witnessing a change of the kind of atomic polarization. It is well-known
that optical pumping using circularly-polarized light of states with
$J>1/2$ also creates alignment along the light propagation \citep{castagna_measurement_2011}.
At low ellipticity, the signal is similar to the Hanle resonance of
an oriented spin-1/2, which HWHM is $\Gamma$, twice the one of an
aligned spin-1. When $\varphi$ increases, the HWHM decreases due
to the contribution of the alignment longitudinal\footnote{The alignment contribution in Eq. \ref{eq:kappaphi}, proportional
to $m_{0}^{(2)}$ and $m_{\pm2}^{(2)}$, has an amplitude varying
as $\sin^{4}(\varphi)/2$ and its HWHM is $\Gamma/2$. Similarly the
orientation contribution, proportional to $m_{0}^{(1)}$, scales as
$\sin^{2}(2\varphi)/2$, and its HWHM is $\Gamma$. As $\varphi$
increases, the alignment contribution to the signal becomes more significant,
thus reducing the amplitude and HWHM of the resonance.} to $\overrightarrow{z}$ . In other words, we observe the sum of
the Hanle effect signals of an oriented spin-1/2 and an aligned spin-1.
\item The $B_{x}$ resonance amplitude does not cancel for any $\varphi$.
Indeed, since this component is orthogonal to both orientation and
alignment, $B_{x}$ is always transverse to the atomic polarization.
When $\varphi<40{^\circ}$, both alignment along $\overrightarrow{y}$
and orientation along $\overrightarrow{z}$ (and some alignment along
$\overrightarrow{z}$) contribute to the Hanle resonance signal. The
amplitude of the signal increases with $\varphi$, reaching its maximum
at $\varphi=\tan^{-1}(1/\sqrt{2})\approx35.2{^\circ}$. For $\varphi>40{^\circ}$,
the alignment along $\overrightarrow{y}$ becomes smaller and orientation
along $\overrightarrow{z}$ keeps increasing, yielding similar resonances
as with $B_{y}$. The data of Fig. \ref{fig:4}.b comforts this interpretation,
showing that the $B_{x}$ resonance HWHM goes from the one of an aligned
state Hanle resonance to the one of a spin-1 pumped with circularly-polarized
light.
\end{itemize}
Although the shape of the dependence is qualitatively good, there
is not a good quantitative agreement between the theoretical predictions
and the measurements. The Fig. \ref{fig:4}.a show a good agreement
for high ellipticities which worsens for $\varphi<30{^\circ}$. Further
analysis suggests that the discrepancies come from the isotropic nature
of the relaxation rate $\Gamma$ used as hypothesis. Through the method
of Appendix \ref{sec:AppD}, this rate can be decomposed on $\Gamma_{e}$
and $\Gamma_{p}$ as shown in Fig. \ref{fig:5}. As expected, the
optical pumping rate $\Gamma_{p}$ does not vary with $\varphi$ because
of the low enough optical power used. The relaxation rate $\Gamma_{e}$
decreases with $\varphi$, witnessing a spin-dependent relaxation
process of unknown nature.

We probed if including explicitly a spin-dependent relaxation in the
model could improve the fit with experimental data. We calculated
the resonance signals with anisotropic $\Gamma$ by solving Eq. \ref{eq:3step}
with $\Gamma^{(k)}=\Gamma_{p}+\Gamma_{e}^{(k)}$. The expressions
are cumbersome and not given here. The fit is made in several steps:
first $\Gamma_{e}^{(2)}$ is fitted from the HWHM variation with $\varphi$
for the $B_{z}$ resonances. Then $\Gamma_{e}^{(1)}$ is fitted from
the HWHM variation with $\varphi$ for the $B_{x}$ and $B_{y}$ resonances.
The result (Fig. \ref{fig:4}, dashed lines) shows a much better agreement
with the measurements.

The nature of this spin-dependent relaxation is unclear. A well-known
spin-dependent relaxation process of the $2^{3}\mathrm{S}_{1}$ state
in $^{4}$He plasma is Penning ionization, which is inhibited when
all the atoms are prepared in the $\left|+1\right\rangle $ or the
$\left|-1\right\rangle $ state. However, according to Ref. \citep{mccusker_intense_1969}
the steady state electron density resulting from Penning ionization
is proportional to $n_{0}^{2}+2(n_{1}n_{0}+n_{1}n_{-1}+n_{0}n_{-1})$,
where $n_{i}$ is the state population of the $2^{3}\mathrm{S}_{1}$
state Zeeman sub-level with $m_{z}=i$. This leads to a higher relaxation
rate for metastable $^{4}$He atoms pumped with circularly-polarized
light ($n_{\pm1}=0$, $n_{0,\mp1}\neq0$) than for linearly-polarized
light ($n_{0}=0$). We thus believe that other collisional processes
in the helium plasma, maybe involving impurities, could be responsible
for this unexpected behaviour. A similar behaviour related to impurities
has been reported long ago with optically-pumped mercury \citep{barrat_depolarisation_1966}.

\section{Single-RF parametric resonance of a spin-1 state pumped with elliptically-polarized
light}

\label{sec:Single-RF}
\begin{figure*}[t]
\centering

\includegraphics[scale=0.95]{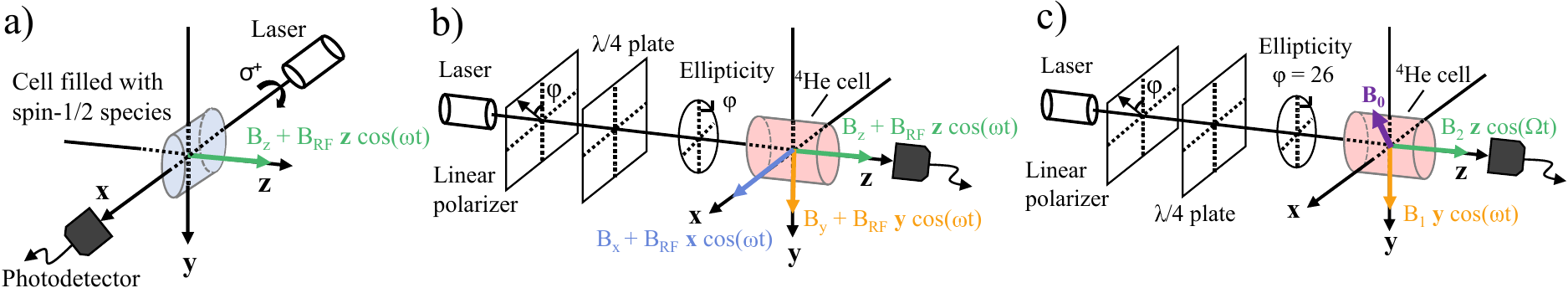}

\caption{\label{fig:6}Schemes of the different PRMs geometries considered.
(a) Scheme for a single-RF PRM using an oriented spin-1/2 atomic species.
(b) Geometry considered in Sec. \ref{sec:Single-RF} for the study
of single-RF parametric resonance in $^{4}$He as a function of $\varphi$.
Only one component of the magnetic field and its parallel RF field
are non-zero at once. (c) Geometry of the $^{4}$He elliptically-polarized
light based PRM. $\protect\overrightarrow{B_{0}}$ is the static magnetic
field to be measured.}
\end{figure*}

Let us briefly recall the zero-field parametric
resonance phenomenon. For a spin-1/2 state, optically pumped with
circularly-polarized light propagating along $\overrightarrow{x}$
(Fig. \ref{fig:6}.a), a transverse oscillating RF magnetic field
$B_{RF}\overrightarrow{z}\cos(\omega t)$ yields oscillating components
in the photodetection signal at $\omega$ and its harmonics. When
sweeping $B_{z}$ around the null field, the odd harmonics components
show an odd-symmetric Lorentzian dependence with respect to $B_{z}$.
Such a resonance can be observed with respect to the $B_{y}$ component
if the RF-field is applied along the $\overrightarrow{y}$ axis, but
not with the $B_{x}$ component.

For a spin-1 aligned state, a similar resonance is observed when the
applied RF field is transverse to the light polarization direction
\citep{beato_theory_2018}.

Parametric resonances based on circularly- and linearly-polarized
pumping are broadly used to build compact vector OPMs, as they allow
one to measure several components of the magnetic field using only
one optical beam \citep{shah_fully_2018,colombo_four-channel_2016,fourcault_helium-4_2021}.

We study here parametric resonances when the pumping light is elliptically-polarized.
A scheme of the different geometries we consider is shown in Fig.
\ref{fig:6}.b.

In Sec. \ref{sec:Hanle}, we showed that using elliptically-polarized
pumping light on a spin-1 state leads to Hanle resonances for the
three components of the magnetic field. Thus by applying an oscillating
RF field, one can also observe parametric resonances for the three
components of the magnetic field. We can obtain the one-RF parametric
resonance signals as a function of $\varphi$ by using the dressed
atom formalism \citep{dupont-roc_etude_1971,beato_theory_2018} and
Eq. \ref{eq:3step}, yielding amplitudes of the component at frequency
$\omega$:

\begin{widetext}

\begin{equation}
\kappa_{1RF\parallel\overrightarrow{z},\omega}(\omega_{z},\omega_{x,y}=0)\propto\frac{2\Gamma_{p}J_{0,2}J_{1,2}\cos^{2}(2\varphi)\omega_{z}}{\Gamma^{2}+4\omega_{z}^{2}}\label{eq:kappa1RFBz}
\end{equation}

\begin{equation}
\kappa_{1RF\parallel\overrightarrow{x},\omega}(\omega_{x},\omega_{y,z}=0)\propto2\Gamma_{p}\cos^{2}(\varphi)\left[\frac{2J_{0,1}J_{1,1}\sin^{2}(\varphi)}{\Gamma^{2}+\omega_{x}^{2}}+\frac{J_{0,2}J_{1,2}\cos^{2}(\varphi)}{\Gamma^{2}+4\omega_{x}^{2}}\right]\omega_{x}\label{eq:kappa1RFBx}
\end{equation}

\begin{equation}
\kappa_{1RF\parallel\overrightarrow{y},\omega}(\omega_{y},\omega_{x,z}=0)\propto2\Gamma_{p}\sin^{2}(\varphi)\left[\frac{2J_{0,1}J_{1,1}\cos^{2}(\varphi)}{\Gamma^{2}+\omega_{y}^{2}}+\frac{J_{0,2}J_{1,2}\sin^{2}(\varphi)}{\Gamma^{2}+4\omega_{y}^{2}}\right]\omega_{y}\label{eq:kappa1RFBy}
\end{equation}

\end{widetext}

\noindent where the $J_{n,q}=J_{n}(q\gamma B_{RF}/\omega)$ are the
first kind $n$th-order Bessel functions.

\begin{figure*}[t]
\centering

\includegraphics[scale=0.87]{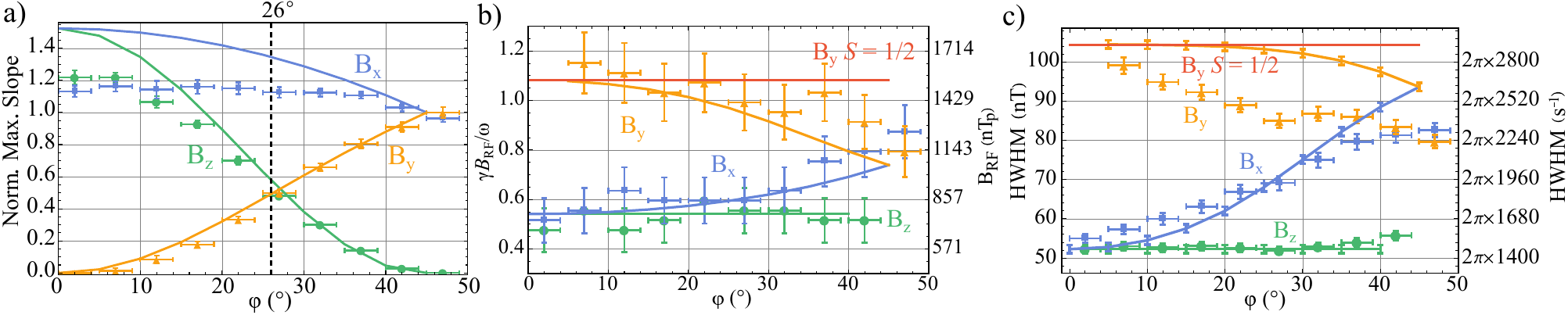}

\caption{\label{fig:7}Study of the single-RF parametric resonance with elliptically-polarized
pumping light. (a) Measured single-RF PRM signals slopes for the three
axes along with the theoretical estimations as a function of $\varphi$.
$B_{RF}$ is chosen to maximize the slope. The experimental and theoretical
values are normalized to their respective value for $B_{y}$ at $\varphi=45{^\circ}$.
(b) $B_{RF}$ maximizing the single-RF PRM signals slopes as a function
of $\varphi$. (c) HWHM of the single-RF resonance at the $B_{RF}$
value maximizing the slopes. The theoretical values are calculated
by the very same method as for the Hanle resonances HWHMs (see Appendix
\ref{sec:AppC}). The error bars of the theoretical estimations are
due to the uncertainty on the optimal value of $B_{RF}$. The red
line shows the theoretical value for an oriented spin-1/2 along $\protect\overrightarrow{z}$.}
\end{figure*}

The Fig. \ref{fig:7}.a shows the experimentally measured slopes acquired
at an optical power of approximately $240$ \textmu W at the cell
input. The RF field with $\omega/2\pi=40\:\mathrm{kHz}$ is applied
along the direction of the magnetic field sweep. The photodetection
signal is demodulated at $\omega/2\pi$ using a Zürich MFLI lock-in
amplifier. The slopes are determined by a linear fit around zero field.
The Fig. \ref{fig:7}.b shows the experimentally measured $\gamma B_{RF}/\omega$
ratio maximizing the signal slope as a function of $\varphi$, along
with the theoretical predictions.

We obtain a qualitative agreement with the measurements. The slope
variations with $\varphi$ is close to the one expected from the $a/\Lambda^{2}$
of Hanle resonances (Fig. \ref{fig:4}.c, see Sec. \ref{subsec:Link-Hanle-RP}
for details). The variations of the HWHM and the optimal $B_{RF}$
witness the kind of atomic polarization which evolve in the magnetic
field. For instance for the $B_{z}$ resonance, $\gamma B_{RF}/\omega=0.54$
and it does not vary with $\varphi$, showing that parametric resonance
is associated only to the alignment along $\overrightarrow{y}$. For
the $B_{y}$ resonance the ratio varies with $\varphi$, ranging from
$\gamma B_{RF}/\omega\thickapprox1.1$---the optimum for a spin-1/2
oriented state---at low $\varphi$ to $0.74$ when the light is circularly
polarized. This behavior is interesting: at low light ellipticity,
the parametric resonance is mainly due to the orientation along $\overrightarrow{z}$.
When $\varphi$ increases, so does the alignment along $\overrightarrow{z}$
and the optimum becomes closer to the one of a spin-1 state pumped
with circularly polarized light. Finally for the $B_{x}$ resonance,
the $\gamma B_{RF}/\omega$ ratio varies from $0.54$ to $0.74$,
showing that at low $\varphi$ the parametric resonance is dominated
by the alignment along $\overrightarrow{y}$, and at higher $\varphi$
by the orientation and alignment along $\overrightarrow{z}$.

The Fig. \ref{fig:7}.c showing the resonances HWHM as a function
of $\varphi$ comforts those interpretations. The HWHM is constant
with $\varphi$ for the $B_{z}$ resonance. For $B_{y}$ the HWHM
evolves from the one corresponding to an oriented spin-1/2 towards
the one of a spin-1 pumped with circularly-polarized light. Finally,
for $B_{x}$ the HWHM varies from the one of alignment resonance to
the one of a spin-1 pumped with circularly-polarized light. As for
the Hanle effect signals, the use of an isotropic relaxation rate
on the model importantly contributes to the discrepancies between
the theoretical predictions and the measurements.

\section{Double-RF parametric resonance magnetometer using elliptically-polarized
pumping light}

\label{sec:Double-RF}In this section, we study the dynamics of the
spin-1 ensemble pumped with elliptically-polarized light when two
RF fields are applied.

We first derive general expressions of the signals as a function of
$\varphi$ (Sec. \ref{subsec:Two-RF-PRM_26}). Then we perform an
experimental study focusing only on the optimal ellipticity value,
found in \citep{le_gal_parametric_2021}. As explained below this
value can be inferred from Hanle resonances measurements.

\subsection{Link between the Hanle resonances and the parametric resonances}

\label{subsec:Link-Hanle-RP}Hanle absorption signals (Eqs. \ref{eq:kappaHanleBz},
\ref{eq:kappaHanleBx} and \ref{eq:kappaHanleBy}) do not display
linear dependence with any component of the magnetic field. However,
when an oscillating RF field is applied (Sec. \ref{sec:Single-RF}),
it leads to modulations in the absorbed light, some of which display
such a linear dependence with the component of the field parallel
to the RF field \citep{cohen-tannoudji_diverses_1970,dupont-roc_etude_1971}.

By comparing the absorption coefficient of Hanle and parametric resonances,
for a spin-1/2 state pumped with circularly-polarized light propagating
along $\overrightarrow{z}$, one finds \citep{dupont-roc_etude_1971}:

\begin{multline}
\kappa_{1RF}^{\sigma^{+}}(\omega_{x},\omega_{y,z}=0)=2J_{0,1}J_{1,1}\frac{a\omega_{x}}{\Gamma^{2}+\omega_{x}^{2}}\\
=2J_{0,1}J_{1,1}\dfrac{\omega_{x}}{\Gamma}\kappa_{Hanle}^{\sigma^{+}}(\omega_{x},\omega_{y,z}=0)\label{eq:1RF1/2}
\end{multline}

\noindent where $a$ is a coefficient related to the light properties
only.

The slope $\partial\kappa_{1RF}^{\sigma^{+}}/\partial B_{x}$ is proportional
to the ratio $a/\Gamma^{2}$ from Hanle resonances, which is however
lowered by the $J_{0,1}J_{1,1}$ factor. Thus, studying the dependence
of $a/\Gamma^{2}$ with $\varphi$ yields an estimation of the best
slope that can be reached for each $\varphi$ for a single-RF PRM.
This remains valid when a second RF field is applied, but with a prefactor
comprising a more complicated combination of Bessel functions.

The Fig. \ref{fig:4}.c shows the variations of $a/\Lambda^{2}$ with
$\varphi$ (we here note the fitted HWHM as $\Lambda$ to avoid confusion
with $\Gamma=\Gamma_{e}+\Gamma_{p}$). The variations closely follow
the single-RF slopes shown on Fig. \ref{fig:7}.a, which confirms
that $a/\Lambda^{2}$ is indeed an appropriate figure of merit of
the PRM slope.

This allows us to obtain the optimal ellipticity for a two-RF PRM:
$\varphi=26{^\circ}$, where the slope to $B_{z}$ equals the $B_{y}$
one. At this ellipticity, the slope to $B_{z}$ ($B_{y}$ respectively)
cannot be increased further without degrading the one to $B_{y}$
($B_{z}$ respectively) while the slope to $B_{x}$ is still higher
than the two others.

\subsection{Two-RF PRM with $\varphi=26{^\circ}$}

\label{subsec:Two-RF-PRM_26}We now study the dynamics when two oscillating
RF fields are applied to $^{4}$He metastable atoms pumped using elliptically-polarized
light with $\varphi=26{^\circ}$.

In alignment-based PRMs, the two well-resolved components of the magnetic
field are the ones parallel to the RF fields \citep{beato_theory_2018},
which are applied orthogonally to the pumping direction (light polarization).
This is the same for orientation-based PRMs, except that the pumping
direction is along the light propagation \citep{dupont-roc_etude_1971}.

When pumping with elliptically polarized light, each component of
the magnetic field is orthogonal to either the orientation along $\overrightarrow{z}$
($B_{y}$), the alignment longitudinal to $\overrightarrow{y}$ ($B_{z}$),
or both of them ($B_{x}$). Thus, the RF fields can be applied along
these three directions, while keeping some sensitivity to all components
of the magnetic field. In the previous section, we found that at $\varphi=26{^\circ}$
the slope to $B_{x}$ is larger than the others. Since the dressing
by a RF field reduces the sensitivity of the axes orthogonal to it
\citep{dupont-roc_etude_1971,beato_theory_2018}, the optimal directions
for applying RF fields in our case seem to be the two orthogonal to
$B_{x}$, so that the two less resolved components ($B_{y}$ and $B_{z}$)
are less degraded than the best-resolved one ($B_{x}$).

As shown in Fig. \ref{fig:6}.c, we consider the two oscillating RF
fields $\overrightarrow{B_{1}}=B_{1}\overrightarrow{y}\cos(\omega t)$
and $\overrightarrow{B_{2}}=B_{2}\overrightarrow{z}\cos(\Omega t)$,
with $\omega\gg\Omega\gg\Gamma,\gamma B_{i}$ \citep{dupont-roc_etude_1971}.
The PRM signals can be calculated from Eqs. \ref{eq:3step}, \ref{eq:Mpphi}
in the doubly dressed-atom picture \citep{dupont-roc_etude_1971,beato_theory_2018},
and Eq. \ref{eq:kappaphi}. Keeping only first-order terms in magnetic
field, we obtain for the component modulated at frequency $\Omega$:

\begin{figure*}[t]
\includegraphics[scale=0.75]{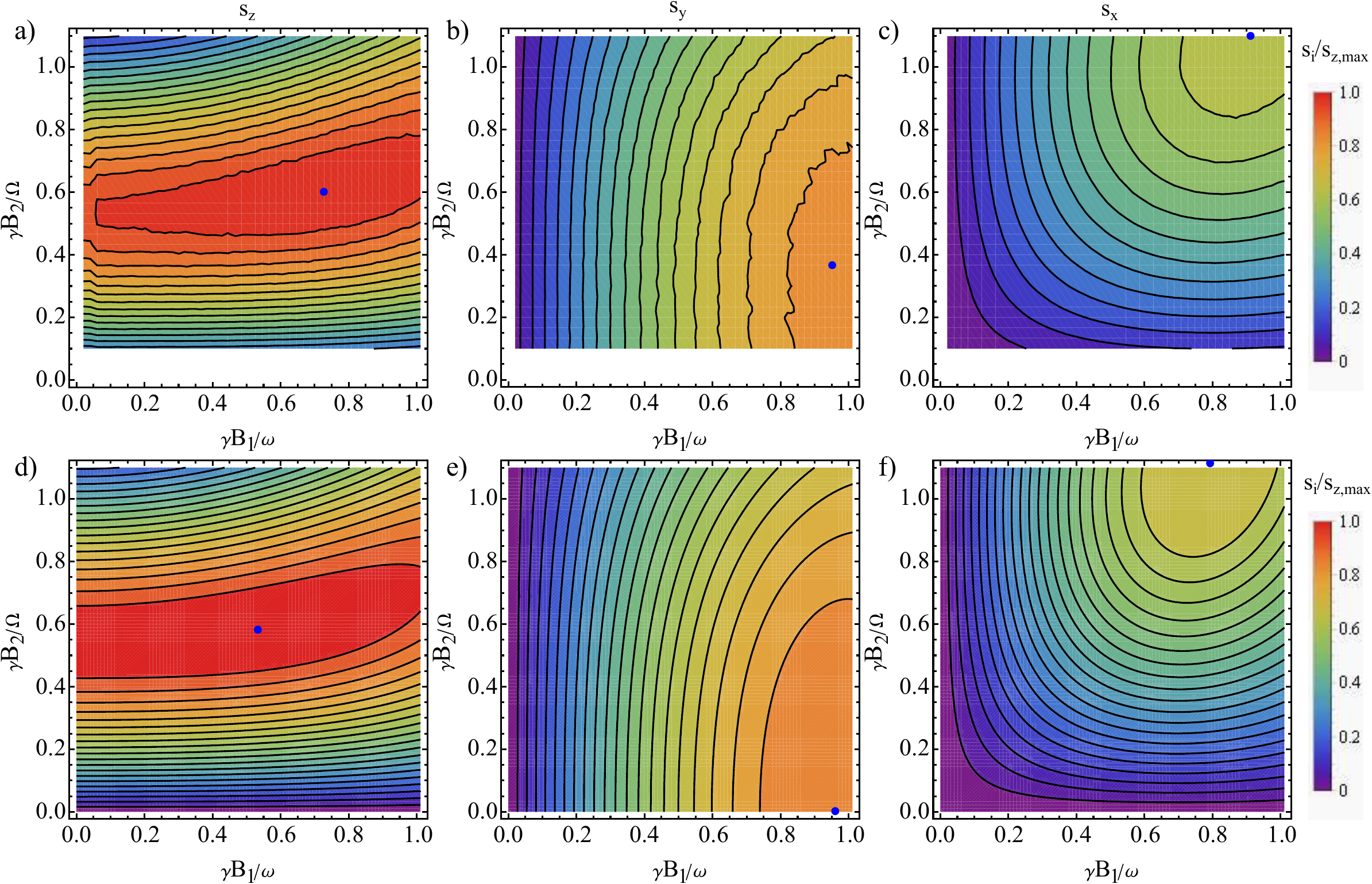}

\caption{\label{fig:8}Slopes of the two-RF PRM based on elliptically-polarized
light. (a), (b) and (c): Experimentally measured slopes $s_{z}$,
$s_{y}$ and $s_{x}$, respectively, as a function of the RF fields
amplitudes, ranging from $B_{1}=58\:\mathrm{nT_{p}}$ to $1444\:\mathrm{nT_{p}}$
($\Leftrightarrow\gamma B_{1}/\omega=0.04$ to $1.01$) for the fast
RF field ($\omega/2\pi=40$ kHz), and from $B_{2}=32.1\:\mathrm{nT_{p}}$
to $353\:\mathrm{nT_{p}}$($\Leftrightarrow\gamma B_{2}/\Omega=0.1$
to $1.1$) for the slow RF field ($\Omega/2\pi=9$ kHz). The blue
dots show the position of the maximum slope for each axis. The three
figures are normalized to the maximum slope reached among the three
axes, $s_{z,max}$, corresponding to the blue dot of Fig. \ref{fig:8}.a.
(d), (e) and (f): Theoretical estimations of $s_{z}$, $s_{y}$ and
$s_{x}$, respectively. The blue dots show the position of the maximum
slope for each axis. The three figures are normalized with the maximum
slope computed among the three axes, $s_{z,max}$ (blue dot on Fig.
\ref{fig:8}.d).}
\end{figure*}

\begin{widetext}

\begin{equation}
\kappa_{\Omega}(\omega_{z},\omega_{x}=\omega_{y}=0)\propto\frac{\Gamma_{p}J_{0,1}[1+3\cos(2\varphi)-2J_{0,2}\sin^{2}(\varphi)]^{2}\mathscr{J}_{0,2}\mathscr{J}_{1,2}}{8\Gamma^{2}}\omega_{z}+O(\omega_{z}^{2})=s_{z}\omega_{z}+O(\omega_{z}^{2})\label{eq:kappaOmega}
\end{equation}

For the component at frequency $\omega$:

\begin{multline}
\kappa_{\omega}(\omega_{y},\omega_{x}=\omega_{z}=0)\propto\frac{\Gamma_{p}\mathscr{J}_{0,1}^{2}\sin^{2}(\varphi)}{4\Gamma^{2}}\left\{ 16J_{0,1}J_{1,1}\cos^{2}(\varphi)-J_{1,2}\left[(\mathscr{J}_{0,2}-1)(1+3\cos(2\varphi)\right.\right.\\
\left.\left.-2J_{0,2}(3+\mathscr{J}_{0,2})\sin^{2}(\varphi)\right]\right\} \omega_{y}+O(\omega_{y}^{2})=s_{y}\omega_{y}+O(\omega_{y}^{2})\label{eq:kappaomega}
\end{multline}

And for their first inter-harmonic at $\omega\pm\Omega$:

\begin{multline}
\kappa_{\omega\pm\Omega}(\omega_{x},\omega_{y}=\omega_{z}=0)\propto\frac{\Gamma_{p}J_{0,1}\mathscr{J}_{0,1}\mathscr{J}_{1,1}}{4\Gamma^{2}}\left\{ \left[8J_{0,1}J_{1,1}\sin^{2}(2\varphi)+2J_{1,2}\sin^{2}(\varphi)\left[(2J_{0,2}(3-\mathscr{J}_{0,2})\sin^{2}(\varphi)\right.\right.\right.\\
\left.\left.\left.+(1+\mathscr{J}_{0,2})(1+3\cos(2\varphi))\right]\right]\right\} \omega_{x}+O(\omega_{x}^{2})=s_{x}\omega_{x}+O(\omega_{x}^{2})\label{eq:kappaOmom}
\end{multline}

\end{widetext}

\noindent where $J_{n,q}=J_{n}(q\gamma B_{1}/\omega)$ and $\mathscr{J}_{n,q}=J_{n}(qJ_{0,1}\gamma B_{2}/\Omega$).

The Fig. \ref{fig:8}.a, b and c show the slopes measured for each
component of the magnetic field as a function of the two RF fields
amplitudes. The RF fields frequencies are $\omega/2\pi=40$ kHz and
$\Omega/2\pi=9$ kHz, the optical power is set to approximately $250$
\textmu W at cell input and we apply magnetic fields sweeps of $\pm90$
nT at 1 Hz frequency for the three components. The photodetection
signal is then demodulated using a Zürich MFLI lock-in amplifier with
reference signals at $\omega$, $\Omega$ and $\omega\pm\Omega$ for
$B_{y}$, $B_{z}$ and $B_{x}$ respectively. These measurements can
be compared to the theoretical predictions (Eqs. \ref{eq:kappaOmega},
\ref{eq:kappaomega} and \ref{eq:kappaOmom}) shown in Figs. \ref{fig:8}.d,
e and f.

There is a good agreement between the experiment and the theoretical
expectations. A simple physical interpretation of this dynamics is
not straightforward. The dressed-atom formalism shows that the dynamics
is close to the one of Hanle effect, but in a generalized rotating
frame \citep{dupont-roc_etude_1971}. We attempt here to give an interpretation
of the observed behavior in this framework.

The slope $s_{y}$, is maximum where $B_{2}\approx0$. The dominant
contribution is due to the orientation along $\overrightarrow{z}$.
$s_{y}$ depends on the RF field along $\overrightarrow{z}$ with
a $\mathscr{J}_{0,1}^{2}$ prefactor, which has two origins: the dressing
of $B_{y}$ by $\overrightarrow{B_{2}}$, and the $\overline{m}_{\pm1}^{(k)}$
evolution\footnote{The $\overline{m}_{q}^{(k)}$ refer to the dressed atomic multipole
moments before applying the rotation to come back to the laboratory
frame and express the signal, for details see \citep{dupont-roc_etude_1971,beato_theory_2018}.} in $\overrightarrow{B_{2}}$ . Both contributions reduce the slope,
when $\mathscr{J}_{0,1}^{2}\neq1$, i.e. when $B_{2}\neq0$.

The slope $s_{z}$ reaches the highest values among the three axes.
It is enhanced by $\overrightarrow{B_{1}}$. In Eq. \ref{eq:kappaOmega},
$s_{z}$ depends on $B_{1}$ with $J_{0,1}$ and $J_{0,2}$ factors.
The first one lowers $s_{z}$ when $B_{1}\neq0$. Since $\varphi=26{^\circ}$,
if $J_{0,2}=1$ (i.e. $B_{1}=0$), $s_{z}$ decreases due to the $[1+3\cos(2\varphi)-2J_{0,2}\sin^{2}(\varphi)]^{2}$
factor in Eq. \ref{eq:kappaOmega}. The compromise between the $J_{0,1}$
and $J_{0,2}$ contributions leads to an optimal $s_{z}$ when $B_{1}\neq0$,
so that $J_{0,2}<1$.

\begin{figure*}[t]
\includegraphics[scale=0.8]{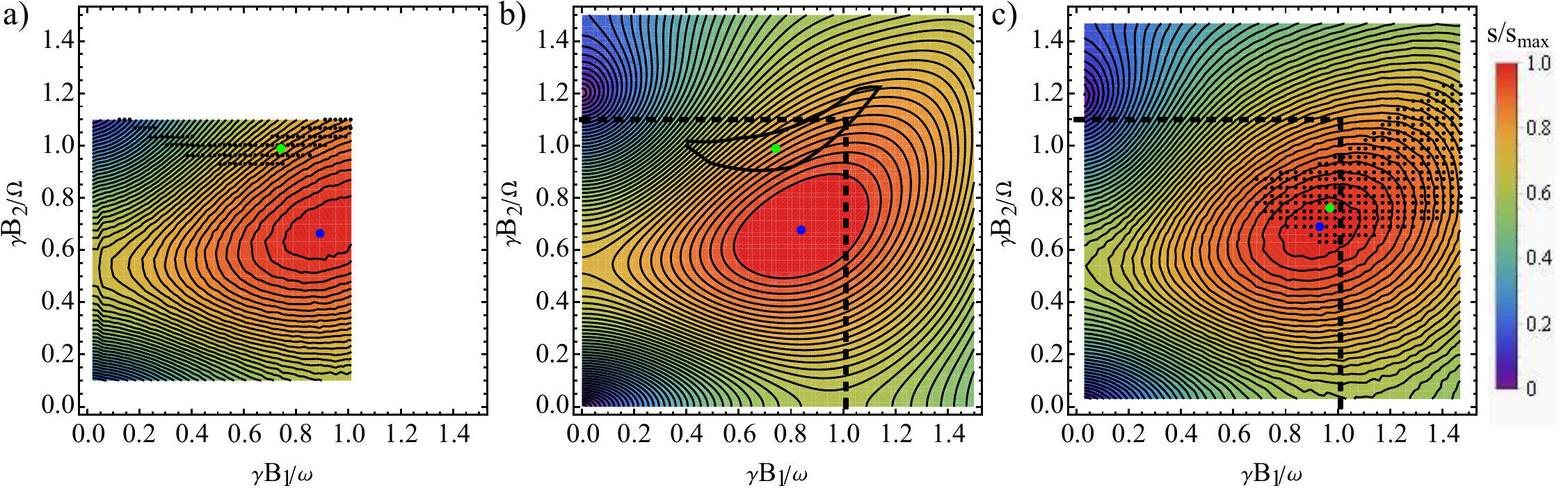}

\caption{\label{fig:9}(a) Experimental dependence of $s$ with the RF fields
amplitudes, for $\omega/2\pi=40$ kHz and $\Omega/2\pi=9$ kHz. The
values are normalized to the maximum value $s_{max}$ (blue dot which
coordinates are $\gamma B_{1}/\omega=0.89$, $\gamma B_{2}/\Omega=0.67$).
The green dot is where $s_{x}\approx s_{y}\approx s_{z}$, and has
coordinates $\gamma B_{1}/\omega=0.74$, $\gamma B_{2}/\Omega=0.99$.
The black dotted area is where the isotropic condition Eq. \ref{eq:isotropcond}
is fulfilled. (b) Theoretical estimations of $s$ with the RF fields
amplitudes. The values are normalized with the maximum value $s_{max}$
(blue dot which coordinates are $\gamma B_{1}/\omega=0.84$, $\gamma B_{2}/\Omega=0.68$).
The green dot is where $s_{x}=s_{y}=s_{z}$, and has coordinates $\gamma B_{1}/\omega=0.74$,
$\gamma B_{2}/\Omega=0.99$. The solid black contoured area is where
the isotropic condition is fulfilled. The black dashed square delimits
the area of Fig. \ref{fig:9}.a. (c) Experimental dependence of $s$
with the RF fields amplitudes, for $\omega/2\pi=40$ kHz and $\Omega/2\pi=15$
kHz. The values are normalized with the maximum value $s_{max}$ (blue
dot which coordinates are $\gamma B_{1}/\omega=0.93$, $\gamma B_{2}/\Omega=0.69$).
The green dot shows the RF amplitudes for which $s_{x}\approx s_{y}\approx s_{z}$
(i.e. $I_{x}\approx I_{y}\approx I_{z}\approx0.33$, see Appendix
\ref{sec:AppE}), which has coordinates $\gamma B_{1}/\omega=0.97$,
$\gamma B_{2}/\Omega=0.76$. The black dotted area is where the isotropic
condition is fulfilled. The black dashed square delimits the area
of Fig. \ref{fig:9}.a.}
\end{figure*}

Finally, there is a linear dependence at the first inter-harmonic
$\omega\pm\Omega$ with $B_{x}$. The slope $s_{x}$ comes from the
doubly-dressed atomic multipole moments bearing the linear dependence
with $B_{x}$ ($\overline{\overline{m}}_{\pm1}^{(1),(2)}$). They
are modulated once at the frequency of each RF field when coming back
to the laboratory frame. At $\varphi=26{^\circ}$, $s_{x}$ strongly
benefits from both orientation and alignment, as expected from the
Hanle effect measurements (Sec. \ref{sec:Hanle}). This is the main
origin of the slope increase for this axis---without parallel RF
field---compared to the usual alignment- or orientation-based PRMs,
allowing to reach isotropy with reasonable slope degradation \citep{le_gal_parametric_2021}.

It is finally interesting to discuss for which parameters such PRM
scheme allows to obtain isotropic sensitivities (i.e. $s_{x}\approx s_{y}\approx s_{z}$),
as discussed in the reference \citep{le_gal_parametric_2021}. The
Fig. \ref{fig:9}.a shows $s=\sqrt{s_{x}^{2}+s_{y}^{2}+s_{z}^{2}}$
obtained from the experimental data presented in Figs. \ref{fig:8}.a,
b and c, along with the theoretical prediction for $s$ (Fig. \ref{fig:9}.b),
and the experimentally measured $s$ when $\Omega/2\pi=15\:\mathrm{kHz}$
instead of $9\:\mathrm{kHz}$ (Fig. \ref{fig:9}.c). The dotted area
on the three figures shows the RF amplitudes values for which the
``isotropic condition'' (Eq. \ref{eq:isotropcond}), as presented
in Appendix \ref{sec:AppE}, is fulfilled.

The agreement between the theoretical predictions and the measurement
at $\Omega/2\pi=9$ kHz is good, both for the values of $s$ and for
the region of isotropy. However, the ($B_{1}$, $B_{2}$) regions
for which the isotropic sensitivity condition is fulfilled do not
overlap the ones where $s$ is maximum. Surprisingly we found that
increasing $\Omega/2\pi$ to $15\:\mathrm{kHz}$ allows obtaining
this overlap \citep{le_gal_parametric_2021}. As shown by the green
dot in Fig. \ref{fig:9}.c, the RF amplitudes leading to optimal isotropic
slopes are $B_{1}=1385\:\mathrm{nT_{p}}$ ($\gamma B_{1}/\omega=0.97$)
and $B_{2}=407\:\mathrm{nT_{p}}$ ($\gamma B_{2}/\Omega=0.76$). At
these RF amplitudes, this is obtained mostly thanks to an increase
of $s_{x}$ when $\Omega/2\pi=15\:\mathrm{kHz}$ compared to $9\:\mathrm{kHz}$,
and leads to a higher absolute value of $s$ in addition of bringing
the isotropic area where $s$ is maximum.

The Figs. \ref{fig:9}.b and c show that the theoretical predictions
are not anymore in good agreement with the measurements, mostly concerning
the isotropic area. Moreover, the theoretical model does not explain
the observed increase of $s_{x}$ with a larger $\Omega$.

\begin{figure*}[t]
\includegraphics{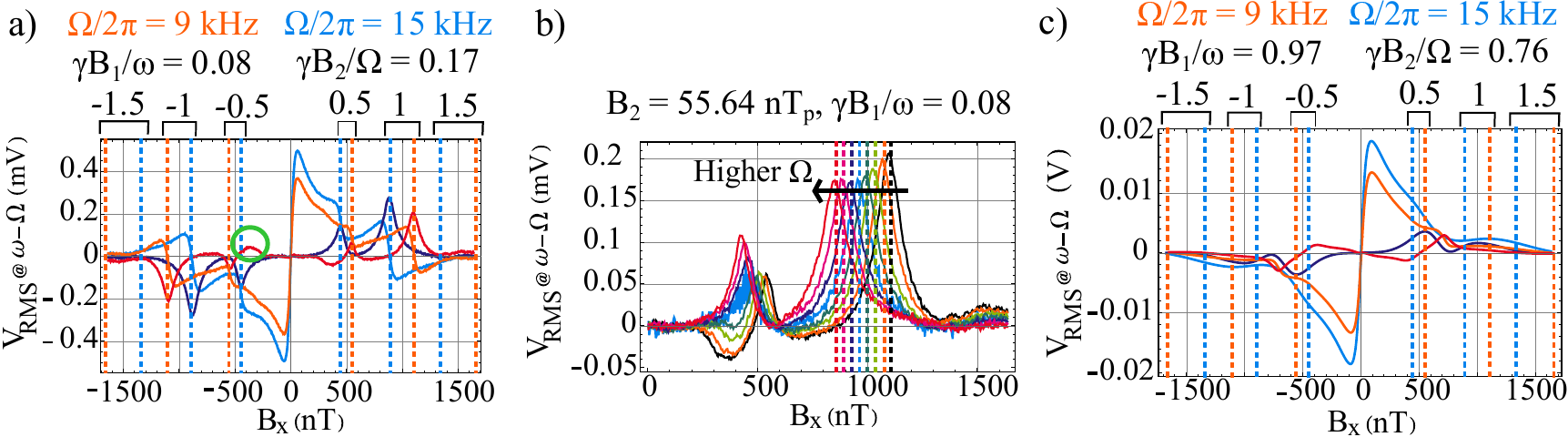}

\caption{\label{fig:10} Resonances in the signal demodulated at $\omega-\Omega$
for a large span $B_{x}$ scanning with $\omega/2\pi=40\:\mathrm{kHz}$.
(a) Resonances at low RF fields amplitudes. The unknown resonance
appearing for $\Omega/2\pi=9\:\mathrm{kHz}$ at $B_{x}<(\omega-\Omega)/(2\gamma)$
is circled in green. (b) In-quadrature demodulated signal for different
values of $\Omega$: $\Omega/2\pi=9\:\mathrm{kHz}$ (black), $9.8\:\mathrm{kHz}$
(orange), $10.5\:\mathrm{kHz}$ (yellow), $11\:\mathrm{kHz}$ (light
green), $12\:\mathrm{kHz}$ (deep green), $13\:\mathrm{kHz}$ (light
blue), $14\:\mathrm{kHz}$ (deep blue), $15\:\mathrm{kHz}$ (pink),
and $16\:\mathrm{kHz}$ (red). The vertical dashed lines show the
value of $(\omega-\Omega)/\gamma$. (c) Resonances at the RF fields
amplitudes of the green dot in Fig \ref{fig:9}.c. For (a) and (c),
the vertical dashed lines show the multiples $n(\omega-\Omega)/\gamma$
for $\Omega/2\pi=9\:\mathrm{kHz}$ in orange and $\Omega/2\pi=15\:\mathrm{kHz}$
in blue, and the orange (light blue) and red (deep blue) lines show
the in-phase and in-quadrature demodulated signals for $\Omega/2\pi=9\:\mathrm{kHz}$
($\Omega/2\pi=15\:\mathrm{kHz}$) respectively.}
\end{figure*}

A further experimental study shows that this increase seems to come
from an influence of resonances adjacent to the zero-field one, which
lie at $\gamma B_{x}=(\omega-\Omega)/2$, being therefore closer to
$B_{x}=0$ when $\Omega/2\pi=15\:\mathrm{kHz}$ as shown in Fig \ref{fig:10}.a.
When $\Omega$ is so that $2\Omega<\omega-\Omega$, there is a new
resonance for $\gamma B_{x}<(\omega-\Omega)/2$, which is absent when
$2\Omega>\omega-\Omega$ at least at low RF amplitudes (Fig \ref{fig:10}.b).
This resonance seems to separate the zero-field resonance from the
one at $\gamma B_{x}=(\omega-\Omega)/2$. At the RF fields amplitudes
of the green dot in Fig \ref{fig:9}.c, all these resonances are broadened
and shifted towards higher values of $B_{x}$. For $\Omega/2\pi=15\:\mathrm{kHz}$,
they are less shifted and other resonances of unknown nature are visible
between $B_{x}=0$ and $\gamma B_{x}=(\omega-\Omega)/2$. These broadened
and shifted adjacent resonances seem to be beneficial for $s_{x}$
when $\Omega/2\pi=15\:\mathrm{kHz}$, whereas they lie at larger values
of $B_{x}$ when $\Omega/2\pi=9\:\mathrm{kHz}$, and thus have less
influence on the zero-field resonance (Fig \ref{fig:10}.c). If these
resonances are broadened when the coupling increases, as magnetic
resonances are, the resonances observed when $\Omega/2\pi=9\:\mathrm{kHz}$
would be less broadened than when $\Omega/2\pi=15\:\mathrm{kHz}$
because $B_{2}$ is lower in the former case in order to have $\gamma B_{2}/\Omega=0.76$.

A more thorough understanding of these influences require supplementary
experimental characterizations as well as a refinement of the theory
presented here, accounting for the resonances other than the zero-field
one.

\section{Conclusion}

In conclusion, we presented here how to compute in the three-step
approach formalism the Hanle resonance signals for any closed $J=1\rightarrow J'=0$
optical transition excited with elliptically-polarized light. The
obtained expressions are in qualitative good agreement with the presented
measurements. We showed that the difference can be explained from
a dependence of the relaxation rate in the dark with the pumping light
ellipticity, probably coming from some collisional processes in the
helium discharge. The introduction of an anisotropic relaxation rate
for orientation and alignment leads to theoretical results much closer
to the experiments.

PRMs signals can be deduced from Hanle effect signals using the dressed-atom
formalism. We gave the single-RF PRM absorption signals as a function
of the light ellipticity pumping a spin-1 atomic state. We also obtained
the two-RFs PRM absorption signals dependence with the light ellipticity
for a specific choice of the oscillating fields direction, which is
the optimal configuration in order to achieve a PRM with isotropic
sensitivity. These expressions show a good agreement with the experiments
as long as the $\Omega\ll\omega$ approximation is fairly fulfilled,
showing that this model allows understanding the dynamics of spin-1
atoms optically pumped with elliptically-polarized light under several
non-resonant oscillating RF fields.
\begin{acknowledgments}
The authors acknowledge R. Romain for his help improving the manuscript,
L.-L. Rouve, G. Pignol, F. Bertrand, T. Jager, J.-M. Léger, M. Le
Prado and E. Labyt for interesting discussions, and W. Fourcault for
his help building the experimental setup. G. LG. acknowledges CEA-LETI
DSYS Ph.D. funding. This research work was supported by the French
ANR via Carnot funding.
\end{acknowledgments}

\appendix

\section{Expression of the rank $k=1$ magnetic evolution matrix in the ITO
basis}

\noindent \label{sec:AppA}The expression $\mathbb{H}(\overrightarrow{B})$
matrix for the rank $k=1$ in the $\{m_{-1}^{(1)},m_{0}^{(1)},m_{1}^{(1)}\}$
basis is:

\begin{equation}
\mathbb{H}(\overrightarrow{B})=-i\gamma\left(\begin{array}{ccc}
-B_{z} & \dfrac{B_{-}}{\sqrt{2}} & 0\\
\dfrac{B_{+}}{\sqrt{2}} & 0 & \dfrac{B_{-}}{\sqrt{2}}\\
0 & \dfrac{B_{+}}{\sqrt{2}} & B_{z}
\end{array}\right)\label{eq:HBori}
\end{equation}

\noindent where $B_{\pm}=B_{x}\pm iB_{y}$.

\section{Expressions of the Hanle signals for arbitrary polarization of the
pumping light}

\noindent \label{sec:AppB}The steady-state solutions of Eq. \ref{eq:3step}
as a function the light ellipticity $\varphi$ are:

\noindent 
\begin{equation}
\begin{array}{c}
m_{0}^{(1)}(\omega_{z},\omega_{x,y}=0)=\dfrac{\Gamma_{p}\sqrt{2}\sin(2\varphi)}{\Gamma}\\
\\
m_{0}^{(2)}(\omega_{z},\omega_{x,y}=0)=-\dfrac{\Gamma_{p}}{2\sqrt{6}\Gamma}\\
\\
m_{\pm2}^{(2)}(\omega_{z},\omega_{x,y}=0)=\dfrac{\Gamma_{p}\cos(2\varphi)}{4(\Gamma\mp2i\omega_{z})}
\end{array}\label{eq:mkqHanleBz}
\end{equation}

\noindent 
\begin{equation}
\begin{array}{c}
m_{0}^{(1)}(\omega_{y},\omega_{x,z}=0)=\dfrac{\Gamma\Gamma_{p}\sqrt{2}\sin(2\varphi)}{(\Gamma^{2}+\omega_{y}^{2})}\\
\\
m_{0}^{(2)}(\omega_{y},\omega_{x,z}=0)=-\dfrac{\Gamma_{p}\left[\Gamma^{2}+\omega_{y}^{2}(1+3\cos(2\varphi)\right]}{2\sqrt{6}\Gamma(\Gamma^{2}+4\omega_{y}^{2})}\\
\\
m_{\pm2}^{(2)}(\omega_{y},\omega_{x,z}=0)=\dfrac{\Gamma_{p}\left[\Gamma^{2}\cos(2\varphi)+\omega_{y}^{2}(1+3\cos(2\varphi)\right]}{4\Gamma(\Gamma^{2}+4\omega_{y}^{2})}
\end{array}\label{eq:mkqHanleBy}
\end{equation}

\noindent 
\begin{equation}
\begin{array}{c}
\begin{array}{c}
m_{0}^{(1)}(\omega_{x},\omega_{z,y}=0)=\end{array}\dfrac{\Gamma\Gamma_{p}\sqrt{2}\sin(2\varphi)}{(\Gamma^{2}+\omega_{x}^{2})}\\
\\
m_{0}^{(2)}(\omega_{x},\omega_{z,y}=0)=-\dfrac{\Gamma_{p}\left[\Gamma^{2}+\omega_{x}^{2}(-1+3\cos(2\varphi)\right]}{2\sqrt{6}\Gamma(\Gamma^{2}+4\omega_{x}^{2})}\\
\\
m_{\pm2}^{(2)}(\omega_{x},\omega_{z,y}=0)=\dfrac{\Gamma_{p}\left[\Gamma^{2}\cos(2\varphi)+\omega_{x}^{2}(-1+3\cos(2\varphi)\right]}{4\Gamma(\Gamma^{2}+4\omega_{x}^{2})}.
\end{array}\label{eq:mkqHanleBx}
\end{equation}

\section{Expressions of the HWHM of Hanle resonances signals}

\noindent \label{sec:AppC}The HWHMs expressions are evaluated as
$\omega_{x,y,z}$ solution of the equation

\noindent 
\begin{multline}
\left[\kappa_{Hanle}(\omega_{x,y,z}\rightarrow\infty)-\left[\kappa_{Hanle}(\omega_{x,y,z}\rightarrow\infty)\right.\right.\\
\left.\left.-\kappa_{Hanle}(\omega_{x,y,z}=0)\right]/2\right]=\kappa_{Hanle}(\omega_{x,y,z}).\label{eq:eqHWHM}
\end{multline}

\noindent This yields:

\begin{widetext}

\begin{equation}
\begin{array}{c}
\mathrm{HWH}\mathrm{M}_{x}(\varphi)=\dfrac{\Gamma\cos(\varphi)}{2}\sqrt{\dfrac{6\left[3-5\cos(2\varphi)\right]+\sqrt{2\left[1331-1500\cos(2\varphi)+369\cos(4\varphi)\right]}}{7+4\cos(2\varphi)-3\cos(4\varphi)}}\\
\\
\mathrm{HWH}\mathrm{M}_{y}(\varphi)=\dfrac{\Gamma}{2}\sqrt{\dfrac{6\left[3+5\cos(2\varphi)\right]+\sqrt{2\left[1331+1500\cos(2\varphi)+369\cos(4\varphi)\right]}}{5+3\cos(2\varphi)}}\\
\\
\mathrm{HWH}\mathrm{M}_{z}(\varphi)=\dfrac{\Gamma}{2}
\end{array}\label{eq:WidthHanle}
\end{equation}

\end{widetext}

\noindent with $\Gamma=\Gamma_{e}+\Gamma_{p}$ the total relaxation
rate.

\section{Determination of the relaxation rate and optical pumping rate from
Hanle resonances signals}

\noindent \label{sec:AppD}The natural relaxation rate $\Gamma_{e}$
and the optical pumping rate $\Gamma_{p}$ can be estimated from the
experimental Hanle resonances by solving the following system:

\noindent 
\begin{equation}
\left\{ \begin{array}{c}
\dfrac{\kappa_{Hanle}(\omega_{x,y,z}=0)}{\kappa_{Hanle}(\omega_{x,y,z}\rightarrow\infty)}=\dfrac{A(B_{x,y,z,0})}{A(B_{x,y,z}\gg\Gamma)}\\
\\
\mathrm{HWH}\mathrm{M}_{x,y,z}(\varphi)=\Lambda_{x,y,z}(\varphi)
\end{array}\right.\label{eq:systGameGamp}
\end{equation}

\noindent where $A(B_{x,y,z,0})=1-V_{\mathrm{PD}}(B_{x,y,z,0})/V_{\mathrm{PD,OFF}}$
is the minimum absorption measured at the value of the natural offset
field component in the magnetic shield. $A(B_{x,y,z}\gg\Gamma)=1-V_{\mathrm{PD}}(B_{x,y,z}\gg\Gamma)/V_{\mathrm{PD,OFF}}$
is the maximum asymptotic absorption value measured at the maximum
magnetic field sweep value. $\Lambda_{x,y,z}(\varphi)$ is the fitted
HWHM for a given resonance. In the first equation, the ratio $\kappa_{Hanle}(\omega_{x,y,z}=0)/\kappa_{Hanle}(\omega_{x,y,z}\rightarrow\infty)$
takes the following values :

\begin{widetext}

\begin{equation}
\begin{array}{c}
\dfrac{\kappa_{Hanle}(\omega_{z},\omega_{x,y}=0)}{\kappa_{Hanle}(\omega_{z}\rightarrow\infty,\omega_{x,y}=0)}=\dfrac{4\Gamma_{e}}{4\Gamma_{e}+3\Gamma_{p}\cos^{2}(2\varphi)}\\
\\
\dfrac{\kappa_{Hanle}(\omega_{x},\omega_{z,y}=0)}{\kappa_{Hanle}(\omega_{x}\rightarrow\infty,\omega_{z,y}=0)}=\dfrac{4\Gamma_{e}}{4\Gamma_{e}+3\Gamma_{p}\cos^{2}(\varphi)(1+3\sin^{2}(\varphi))}\\
\\
\dfrac{\kappa_{Hanle}(\omega_{y},\omega_{z,x}=0)}{\kappa_{Hanle}(\omega_{y}\rightarrow\infty,\omega_{z,x}=0)}=\dfrac{4\Gamma_{e}}{4\Gamma_{e}+3\Gamma_{p}\sin^{2}(\varphi)(1+3\cos^{2}(\varphi))}.
\end{array}\label{eq:ratioGameGamp}
\end{equation}

\end{widetext}

\section{Isotropic condition}

\noindent \label{sec:AppE}We define the condition of isotropic sensitivity
on the slopes of the two-RF PRM as:

\begin{equation}
\left\{ \begin{array}{c}
0.37>I_{x}>0.3\\
\&\\
0.37>I_{y}>0.3\\
\&\\
0.37>I_{z}>0.3
\end{array}\right.\label{eq:isotropcond}
\end{equation}

\noindent with $I_{i}=\left|s_{i}\right|/\left(\left|s_{x}\right|+\left|s_{y}\right|+\left|s_{z}\right|\right)$
where $i\in\{x,y,z\}$. It is chosen so that the slope to each axis
lies in $\pm10\%$ of $1/3$ of the total slope.

\bibliographystyle{apsrev4-2}

%

\end{document}